\documentclass[epsf,prb,twocolumn,showpacs,nofootinbib]{revtex4-1}

\usepackage[pdftex]{graphicx}
\usepackage{dcolumn}
\usepackage{bm}
\usepackage{epsfig}
\usepackage{latexsym}
\usepackage{amsmath}
\usepackage{amsfonts}
\usepackage{amssymb}
\usepackage{color}
\usepackage{array}
\usepackage{framed}

\begin{document}

\title{Finite-Temperature Screening of $U(1)$ Fractons}
\author{Michael Pretko\\
\emph{Department of Physics, Massachusetts Institute of Technology,
Cambridge, MA 02139, USA}}
\date{August 25, 2017}

\begin{abstract}
We investigate the finite-temperature screening behavior of three-dimensional $U(1)$ spin liquid phases with fracton excitations.  Several features are shared with the conventional $U(1)$ spin liquid.  The system can exhibit spin liquid physics over macroscopic length scales at low temperatures, but screening effects eventually lead to a smooth finite-temperature crossover to a trivial phase at sufficiently large distances.  However, unlike more conventional $U(1)$ spin liquids, we find that complete low-temperature screening of fractons requires not only very large distances, but also very long timescales.  At the longest timescales, a charged disturbance (fracton) will acquire a screening cloud of other fractons, resulting in only short-range correlations in the system.  At intermediate timescales, on the other hand, a fracton can only be partially screened by a cloud of mobile excitations, leaving weak power-law correlations in the system.  Such residual power-law correlations may be a useful diagnostic in an experimental search for $U(1)$ fracton phases.
\end{abstract}
\maketitle

\section{Introduction}

Recent work has established the stability of three-dimensional spin liquid phases described by higher rank tensor gauge theories, such as the $U(1)$ higher spin gauge theories \cite{alex,sub,genem} and the ``generalized lattice gauge theories" of Vijay, Haah, and Fu\cite{fracton1,fracton2}, which are the natural discrete analogue.  The most exotic feature of these higher-rank phases is the fact that the emergent tensor gauge field must be coupled to excitations with subdimensional behavior.  These particles are restricted by gauge invariance to exist only in lower-dimensional subspaces of the full three-dimensional system.  For example, some particles can be restricted to motion only along a one-dimensional line, or within a two-dimensional plane.  As an extreme case of this phenomenon, some types of particles are restricted to a 0-dimensional subspace, $i.e.$ a point, and become totally immobile.  Such an immobile excitation is called a fracton, or a 0-dimensional particle.  Theoretical evidence for these immobile particles has been seen for more than a decade \cite{chamon,bravyi,cast,yoshida,haah,haah2}, but interest in the topic has particularly surged just in the last few years\cite{sub,genem,fracton1,fracton2,williamson,sagarlayer,hanlayer,abhinav,mach,parton,slagle,bowen}, establishing the subject as a new subfield of condensed matter physics.

In this work, our focus will be on a higher rank $U(1)$ phase with fracton excitations.  Previous work on this topic \cite{alex,sub,genem} has already established many of the properties of these fractons, such as their conservation laws and their interactions with the tensor gauge field.  This earlier work focused on the zero-temperature behavior of these phases, analyzing the physics of isolated particles.  But what happens to these phases at finite temperature, when we have a soup of thermal excitations?  In particular, is there a smooth crossover to a trivial phase, as in the conventional $U(1)$ spin liquid, or could there be a robust finite-temperature phase?  And to what extent are the finite-temperature properties controlled by the physics of the zero-temperature quantum spin liquid?  Work on the finite-temperature behavior of the discrete fracton models\cite{chamon, abhinav} has already yielded interesting physics, such as long thermalization times and glassy dynamics.  Are there similar stories to be found in the $U(1)$ fracton models?

In the conventional (rank 1) $U(1)$ spin liquid in three dimensions, the finite-temperature behavior is fairly easy to obtain, following from the familiar properties of Maxwell electromagnetism.  At finite temperature, there will be a thermal distribution of the emergent charges which can self-consistently screen out their associated long-range electric fields.  Very close to a charged particle, one will still see the Coulomb field of a point charge, decaying as $1/r^2$.  But at distances longer than some screening length $\lambda$, the electric field will decay exponentially as $e^{-r/\lambda}$, making the particle look effectively neutral ($i.e.$ trivial) at the longest length scales.  Since all excitations are neutralized at long distances, this allows for a smooth crossover at finite temperature between the $U(1)$ spin liquid phase and the trivial phase, where all excitations are neutral.  In the thermodynamic limit, the $U(1)$ spin liquid is therefore only strictly distinct from the trivial phase at zero temperature.

Technically, at any finite temperature, emergent $U(1)$ charges are screened out if one looks at sufficiently long length scales.  But, as we will review, the screening length grows exponentially as the temperature is decreased, $\lambda \propto e^{m/T}$, where $m$ is the mass scale of the charges.  This rapid growth of the screening length allows for spin liquid physics to hold on macroscopic length scales at accessible temperatures.  For example, suppose we are in a finite system of linear size $L$.  Once the temperature drops below $T \sim m/\log L$, the screening length will be of the same order as the system size, so screening will no longer play any significant role in the system.  Since this temperature only decreases logarithmically with system size, it is not unrealistically low, even for macroscopic systems.  Therefore, at low (but accessible) temperatures, the emergent charges are well-defined over macroscopic length scales, and the zero-temperature physics of the $U(1)$ spin liquid dominates the behavior of the system.

In this work, we will generalize the notion of finite-temperature screening to a higher rank $U(1)$ spin liquid phase, finding some other interesting finite-temperature physics along the way.  We will put particular emphasis on a specific phase which hosts both fracton excitations and nontrivial mobile bound states.  We will provide a review of the essential physics of such a phase of matter.  The finite-temperature behavior of this system should be fairly representative of all $U(1)$ phases which possess both fractons and mobile bound states.  This is in contrast to two other types of higher rank $U(1)$ phases: those without fractons, and those without nontrivial mobile bound states.  Such phases have comparatively simpler finite-temperature physics, and we will comment on them towards the end.  For the majority of this work, however, we specialize to the case of $U(1)$ phases which have both fractons and mobile bound states.

We will find that the fracton excitations in such a theory will indeed be screened at finite temperature.  Therefore, this phase has a smooth finite-temperature crossover to the trivial phase, just like the conventional $U(1)$ spin liquid.  However, unlike the normal $U(1)$ spin liquid, there are two distinct temporal regimes associated with screening of a charged disturbance in the system.  At short timescales, the dominant screening mechanism is from thermally excited mobile bound states.  In this regime, which we dub the weak-screening regime, the electric field of a fracton is weaker than the zero-temperature behavior, but still decays as a power law.  At much longer timescales, a fracton can actually be screened by other fractons.  In this strong-screening regime, the electric field of a point charge decays exponentially, and all correlations become short-ranged.

In either case, the composite object of a fracton plus its screening cloud forms a mobile object which can move around the system, albeit at very slow speeds.  In this sense, fracton immobility is technically absent at finite temperature.  However, once again we will find that the screening length grows exponentially as the temperature is lowered.  At sufficiently low temperatures, the zero-temperature physics will dominate on macroscopic length scales, and fractons will be effectively immobile.  Furthermore, even above this temperature, the zero-temperature fracton physics will make its presence felt in the form of a large thermalization time.  This obstruction to thermalization is particularly interesting since it occurs in a completely clean system, without the need for disorder, as in more conventional many-body localization situations.  These unusual finite-temperature properties, such as slow thermalization in the clean limit and persistent power-law correlations, may serve as useful diagnostics in the experimental search for $U(1)$ fracton phases.

\section{The Conventional $U(1)$ Spin Liquid: A Review}

We begin by briefly reviewing a simple calculation for screening in the conventional three-dimensional $U(1)$ spin liquid, which closely follows the treatment of standard electromagnetic screening.  The calculation will be useful in order to compare and contrast with the higher rank case.  Note that the analysis will be almost purely classical.  The particles have a finite energy gap, so at low temperatures the particles will be dilute, making the effects of particle statistics unimportant.  A classical analysis will capture the correct qualitative physics throughout.

We consider a $U(1)$ spin liquid, which is described by an emergent $U(1)$ gauge theory, at a finite temperature $T$.  We take the emergent charges of the $U(1)$ gauge field, both positive and negative, to have some finite mass $m$.\cite{foot1}  At temperature $T$, the equilibrium density of positive and negative charges will be:
\begin{equation}
n_+ = n_- \equiv n_0(T) \propto e^{-m/T}
\end{equation}
(We will not consider bound states of higher charge, which are more energetically costly and contribute comparatively less to screening.)  We now want to know how these thermal excitations respond to a ``test charge" introduced to the system.  We imagine adding an extra point charge $Q$ at a particular fixed location, then examining how the system responds to the resulting change in potential from the disturbance.  The presence of a perturbing potential $\phi(r)$ will shift the Boltzmann weight of the thermally excited particles, causing the density of positive (negative) charges to adjust to:
\begin{equation}
n_\pm = n_0 e^{\mp\beta\phi} \approx n_0(1\mp\beta\phi)
\end{equation}
where we have defined $\beta = T^{-1}$.  The linearization is only valid for a small perturbation, breaking down in the immediate vicinity of the point charge, but captures the correct long-distance physics, which is our primary concern.  Adopting a normalization such that the charge of the emergent particles is $q=1$, the net charge density of the thermal excitations is then:
\begin{equation}
\rho = n_+ - n_- \approx -2n_0\beta\phi
\end{equation}
We recall that, in normal electrostatics, the potential of an arbitrary charge configuration can be written as:
\begin{equation}
\phi(r) = \int d^3r' \frac{\rho(r')}{4\pi|r-r'|}
\end{equation}
where the integral is over three-dimensional space, and we have adopted units such that Gauss's law is $\partial_i E^i = \rho$.  Making use of this equation, the potential around the test charge will take the form:
\begin{equation}
\phi(r) = \int d^3r' \frac{-n_0\beta\phi(r')}{2\pi|r-r'|} + \phi_{bare}(r)
\end{equation}
where $\phi_{bare}$ corresponds to the bare potential of the unscreened test charge, $Q/4\pi r$.  Taking the Laplacian of both sides then gives us the Poisson equation:
\begin{equation}
\partial^2\phi = 2 n_0\beta\phi - Q\delta^{(3)}(r)
\end{equation}
where the delta function represents the test charge.  Away from the delta function (which sets boundary conditions at the origin), the potential obeys $\partial^2\phi = 2n_0\beta\phi$, which has an exponentially decaying solution, $\phi\propto e^{-r/\lambda}$, with a screening length given by:
\begin{equation}
\lambda = (2 n_0\beta)^{-1/2} \propto e^{m/2T}
\end{equation}
where we use the fact that $n_0\propto e^{-m/T}$.  We therefore see that, at distances longer than $\lambda$, any charge in the system will be screened, and there is little remnant of the emergent gauge structure.  This allows for a smooth finite-temperature crossover to a trivial phase, which has only neutral excitations.  However, we note that the screening length grows exponentially at low-temperatures, $\lambda\propto e^{m/2T}$.  For temperatures below $T\sim m/\log L$, which is not unreasonably low, the screening length becomes larger than the system size, and the physics of the $U(1)$ spin liquid will hold throughout the system.

\section{The $U(1)$ Fracton Phase}

\subsection{Review of the Model}

We now wish to generalize the preceding treatment of screening to a higher rank $U(1)$ phase possessing both fractons and also nontrivial mobile excitations.  We begin by reviewing the simplest model of this type.  We expect the same qualitative physics should hold for any $U(1)$ model with these basic characteristics.  The model we will work with has been studied in some detail in previous literature\cite{alex,sub,genem,mach}.  We will review the most important features of this phase here, though we refer the reader to these previous works for more details.  In particular, we will make use of some basic results of the generalized tensor electromagnetism, first established in Reference \onlinecite{genem}.

We take the degrees of freedom of our model to be those of a symmetric rank 2 tensor field $A_{ij}$, defined throughout three-dimensional space, which will play the role of the emergent gauge field.  We call the canonically conjugate variable $E_{ij}$, representing a symmetric tensor electric field.  The theory can be defined by specifying a Gauss's law constraint on the electric field, which in turn specifies the gauge transformation.  For the model we will focus on here, the generalized Gauss's law takes the form:
\begin{equation}
\partial_i\partial_j E^{ij} = \rho
\end{equation}
for scalar charge density $\rho$.  We will take the charges of the system to generically have a finite energy gap.\cite{foot2}  In this case, the low-energy sector is free of charges, $\partial_i\partial_j E^{ij} = 0$, resulting in invariance under the following gauge transformation:
\begin{equation}
A_{ij}\rightarrow A_{ij} + \partial_i\partial_j \alpha
\end{equation}
for scalar gauge parameter $\alpha$ with arbitrary spatial dependence.  This gauge invariance then dictates the form of the low-energy Hamiltonian governing the gapless gauge mode of the theory:
\begin{equation}
H = \int d^3r\,\frac{1}{2}(E^{ij}E_{ij} + B^{ij}B_{ij})
\end{equation}
where the gauge-invariant magnetic tensor takes the form $B_{ij} = \epsilon_{iab}\partial^aA^b_{\,\,\,j}$.  This Hamiltonian leads to a linear dispersion for the gapless gauge mode, including both spin-2 and lower spin degrees of freedom.

The gapless gauge mode is interesting in its own right, but here we will mostly be focusing on the physics of charges, as defined by the Gauss's law, $\rho = \partial_i\partial_j E^{ij}$.  These charges are most notable for their conservation laws.  Like in the conventional $U(1)$ spin liquid, this system will obey conservation of charge:
\begin{equation}
\int d^3r\,\rho = \textrm{constant}
\end{equation}
$i.e.$ the total charge in the system remains fixed.  However, the charges of this higher rank $U(1)$ gauge theory also obey a second conservation law:
\begin{equation}
\int d^3r\,\rho\vec{r} = \textrm{constant}
\end{equation}
which reflects the conservation of dipole moment.  This means that, not only must charge be conserved, but any motion of charges must proceed in such a way that the total dipole moment of the system remains fixed.  The consequences of this new conservation law are severe.  An isolated charge cannot move by itself, since this would change the total dipole moment.  An isolated fundamental charge of this theory is locked in place and is therefore a fracton.

While isolated charges of this theory are locked in place, charges can obtain mobility when they come together to form bound states.  For example, a dipolar bound state of a positive and negative charge is free to move around the system in any direction, as long as its dipole moment remains fixed.  Furthermore, such dipoles are nontrivial excitations of the system, since they cannot be directly created from or absorbed into the vacuum, due to the dipolar conservation law.

There are many other types of $U(1)$ fracton models\cite{sub}, with various gauge structures.  For example, most $U(1)$ gauge theories of rank higher than 2 tend to have fracton excitations.  But in all cases, the physics of fractons is essentially driven by the same basic principle: higher moment conservation laws which restrict the motion of charges.

\subsection{Weak-Screening Regime}

Now that we have our model in hand, we want to know how the physics of screening carries over to this theory.  One would like to once again say that thermally excited charges will rearrange themselves in such a way as to screen out long-range interactions.  However, the fundamental charges of the theory are fractons, which are locked in place and cannot rearrange themselves into a screening pattern.  (We will revisit and adjust this idea later.)  But while the fractons cannot easily react to a potential, the dipoles of the system can.  At nonzero temperature, the system will have a finite density of thermally excited mobile dipoles, which will react to the field of a test charge and can contribute to screening.  However, since the dipoles are charge-neutral objects, we expect that they will not be able to screen as effectively as charged particles could.  Indeed, we will find that dipoles can only partially screen charged disturbances.

Just as we earlier focused on screening due to the minimal charges in the conventional $U(1)$ spin liquid, we will here focus on screening due to the minimal dipoles.  As discussed in earlier work\cite{sub,genem}, the dipole moment in this theory is quantized in a lattice-dependent way.  For example, on a cubic lattice system, a dipole is labeled by three integers, $\vec{p} = (n_x,n_y,n_z)$, in appropriate units.  In this case, the minimal dipoles are $\vec{p}= (\pm 1,0,0)$, $(0,\pm 1,0)$, and $(0,0,\pm 1)$.  Such minimal dipoles will provide the dominant contribution to screening, since they are energetically cheapest and will therefore be thermally excited in the greatest numbers, with a density $n_{0,d}(T)\propto e^{-m_d/T}$, where $m_d$ is the mass of the minimal dipoles.  Including the effects of higher strength dipoles should not affect the qualitative physics.

We will now take advantage of some results regarding the generalized electromagnetism of this phase, established in Reference \onlinecite{genem}.  By imposing the conditions of electrostatics, it is not hard to show that a static electric field can be described via a scalar potential formulation, $E_{ij} = \partial_i\partial_j \phi$.  For a point charge $q$, this scalar potential takes the form\cite{foot3}:
\begin{equation}
\phi(r) = -\frac{qr}{8\pi}
\end{equation}
and the corresponding electric field falls off as $E_{ij}\sim 1/r$.

In the presence of such a scalar potential, the potential energy of a dipole will be $-p^j\partial_j\phi$, which shifts the Boltzmann weight.  The density of thermally excited dipoles with dipole moment $p^j$ will then become:
\begin{equation}
n_p = n_{0,d}e^{\beta p^j\partial_j\phi}\approx n_{0,d}(1+\beta p^j\partial_j\phi)
\end{equation} 
The scalar potential associated with a dipole takes the form:
\begin{equation}
\phi_p(r) = \frac{(p\cdot r)}{8\pi r}
\end{equation}
Using this form of the scalar potential, the total potential around a test charge becomes:
\begin{equation}
\phi(r) = \phi_{bare}(r) + \sum_{\{p^j\}} n_{0,d} \int dr' (1+\beta p^i\partial_i\phi(r')) \frac{p^j(r_j-r_j')}{8\pi|r-r'|}
\end{equation}
where the sum is over the possible orientations of the minimal dipoles.  These orientations depend on the underlying lattice, but for any lattice, each orientation will also appear with its negative, so $\sum p^j = 0$.  This leads to the vanishing of the first term in the sum, leaving us with:
\begin{equation}
\phi(r) = \phi_{bare}(r) +  n_{0,d}\beta \int dr' \bigg(\sum_{\{p^j\}} p^ip^j\bigg) \frac{\partial_i\phi(r') (r_j-r_j')}{8\pi|r-r'|}
\end{equation}
For the cubic lattice, the sum in the integrand evaluates to $2\delta^{ij}$.  For other lattice choices, the sum will also be proportional to $\delta^{ij}$, though the coefficient is non-universal, so we write $\sum_p p^ip^j = \alpha\delta^{ij}$.  Using this, our above equation becomes:
\begin{equation}
\begin{split}
\phi(r) = \,\,&\phi_{bare}(r) +  \alpha n_{0,d}\beta \int dr' \frac{\partial^j\phi(r') (r_j-r_j')}{8\pi|r-r'|} = \\
&\phi_{bare}(r) - \alpha n_{0,d}\beta \int dr' \frac{\phi(r')}{4\pi |r-r'|}
\end{split}
\end{equation}
where we have integrated by parts.\cite{foot4}  We then take a Laplacian to yield:
\begin{equation}
\partial^2\phi = \partial^2\phi_{bare} + \alpha n_{0,d}\beta\phi
\label{dipolescreen}
\end{equation}
Let us now take $\phi_{bare}$ to be that of a bare test charge, which has the form $\phi_{bare} = -Qr/8\pi$.  Taking the Laplacian yields $\partial^2\phi_{bare} = -Q/4\pi r$, so Equation \ref{dipolescreen} becomes:
\begin{equation}
\partial^2\phi = -\frac{Q}{4\pi r} + \alpha n_{0,d}\beta \phi
\end{equation}
Fourier transforming, we obtain:
\begin{align}
-k^2\phi(k) = -\frac{Q}{k^2} + \alpha n_{0,d}\beta\phi(k) \\
\phi(k) = \frac{Q}{k^2(k^2+\alpha n_{0,d}\beta)}
\end{align}
At short distances close to the charge (corresponding to large $k$), we will have $\phi(k) \sim k^{-4}$, which gives $\phi(r)\sim r$.  This is as expected, since at short distances we should see the field of a point charge.  At long distances (small $k$), the scaling is adjusted to $\phi(k)\sim k^{-2}$, so $\phi(r) \sim 1/r$.  The crossover occurs around a screening length of:
\begin{equation}
\lambda = (\alpha n_{0,d}\beta)^{-1/2} \propto e^{m_d/2T}
\end{equation}
using the fact that $n_{0,d}\propto e^{-m_d/T}$.  At distances longer than this, the potential behaves (up to an additive constant) as:
\begin{equation}
\phi(r) = \frac{Q\lambda^2}{4\pi r}
\end{equation}
In this regime, the electric field $E_{ij} = \partial_i\partial_j\phi$ will scale as $1/r^3$.  We therefore see that, after accounting for finite-temperature dipole screening, the bare $1/r$ electric field is screened to a $1/r^3$ field.  This reduction by two powers indicates that the first two terms of the generalized multipole expansion vanish.  In other words, the composite object of the fracton plus its screening cloud is not only neutral, but also has zero dipole moment.  Therefore, the screened fracton is carrying no nontrivial quantum numbers and is a trivial excitation.\cite{foot5}

Since all finite-temperature excitations are effectively trivial at long length scales, this higher rank $U(1)$ spin liquid can smoothly cross over to a trivial phase at finite temperature, just like its more conventional cousin.  Nevertheless, the screening length still grows exponentially at low temperatures, $\lambda\propto e^{m_d/2T}$.  Once again, for $T< m_d/\log L$, the screening length is cut off by the system size, and the physics of the higher rank spin liquid will hold throughout the system.  Even above this temperature, the onset of this type of fracton mobility is quite slow.  In order for the screened charge to move, it must drag along its screening cloud, of linear size $\lambda$.  As discussed in Reference \onlinecite{genem}, the effective mass of such a composite object grows exponentially in the separation of particles, which in this case is of order $\lambda$ on average.\cite{foot51}  In other words, the characteristic velocity of the screened excitations will be exponentially small in the screening length:
\begin{equation}
v_{eff}\propto e^{-(\lambda/a)}\propto e^{-e^{m_d/2T}}
\end{equation}
where $a$ is some lattice scale.  At low temperatures, the mobility of a screened fracton decreases doubly exponentially, leading to negligible fracton motion via this mechanism.  However, there is actually a different finite-temperature mechanism which can lead to slightly faster fracton motion, as we discuss next.

\subsection{Absorptive Diffusion}

\begin{figure}[t!]
 \centering
 \includegraphics[scale=0.25]{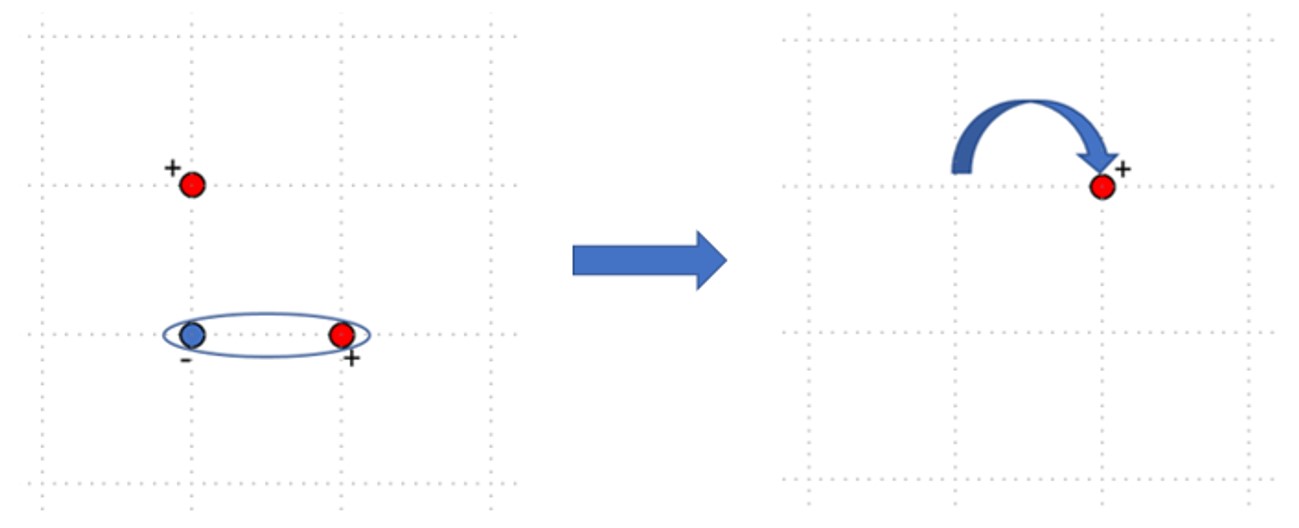}
 \caption{A fracton cannot move by itself.  However, by absorbing a $\hat{p}$-oriented dipole, the fracton can hop a certain distance in the $\hat{p}$ direction.}
 \label{fig:ab}
 \end{figure}

We have found that, once screening is taken into account, a fracton can move around the system, so long as it is willing to drag along its entire screening cloud.  However, such a bulky process is not the most efficient type of fracton motion.  Fractons can move more quickly through the direct absorption of thermally excited dipoles.  As illustrated in Figure \ref{fig:ab}, a fracton can absorb a dipole and thereby hop in the direction of its dipole moment.  In a certain sense, such a dipole is carrying a ``quantum of position," which is transferred to the fracton.  By absorbing randomly directed thermally excited dipoles, the fracton will thereby diffuse around the system, effectively performing a random walk.  We refer to this process as ``absorptive diffusion."

Strictly speaking, in order to conserve momentum, the process depicted in Figure \ref{fig:ab} will have to involve the gapless gauge mode (``graviton").  The initial fracton and the final fracton are both carrying zero momentum, while the dipole momentum is in general nonzero.  The absorption of a dipole must therefore be supplemented by graviton emission in order to conserve momentum and energy.  However, the gravitons are gapless and can be readily emitted and absorbed in a thermal system.  The dominant factor setting the rate of dipole absorption will be the thermal density of dipoles, $n_0$.  The typical velocity of a fracton hopping via this mechanism therefore behaves as:
\begin{equation}
v_{eff}\propto n_{0,d}\propto e^{-m_d/T}
\end{equation}
(The thermal distribution of gravitons, $n_g$, can only provide a power-law prefactor to this exponential scaling.)  This diffusive velocity is must faster than the typical velocity of a screening cloud at low temperatures, which we found in the previous section to behave doubly exponentially in $T$.  Fracton mobility is therefore dominated by this sort of process.  Note, however, that the fracton does not maintain this velocity in one particular direction for a sustained period of time, but rather undergoes diffusive motion with a mean free path set by the lattice scale.

\subsection{Strong-Screening Regime}

At the outset of our finite-temperature analysis, we assumed that the primary screening mechanism comes from thermally excited dipoles, since fractons are largely immobile, and therefore cannot easily rearrange themselves to create a screening cloud.  However, we have now found that there actually are mechanisms by which a fracton can move at finite-temperature, albeit at very slow speeds.  Therefore, we should expect fractons to contribute to screening on sufficiently long time scales.  Before estimating the relevant time scale, let us examine the effects of such fracton-on-fracton screening.

We now assume that the system has had sufficient time for the fracton sector to reach thermal equilibrium, such that we have a thermal distribution of fractons, $n_{0,f}\propto e^{-m_f/T}$, where $m_f$ is the mass scale of fractons.  The scalar potential $\phi$ plays the role of the potential energy per charge associated with fractons.  The presence of such a potential will therefore shift the corresponding Boltzmann weights.  In the presence of a small perturbing potential, the thermal density of fractons becomes:
\begin{equation}
n_\pm =  n_{0,f} e^{\mp\beta\phi}\approx n_{0,f}(1\mp\beta\phi)
\end{equation}
for positive and negative charges, respectively.  The net charge density will then be:
\begin{equation}
\rho = n_+ - n_- = -2n_{0,f}\beta\phi
\end{equation}
As discussed earlier, the potential generated by a unit charge fracton takes the form $\phi = -r/8\pi$.  The net potential around a test charge will then take the form:
\begin{equation}
\phi(r) = \int dr' \frac{n_{0,f}\beta}{4\pi}|r-r'|\phi(r') + \phi_{bare}(r)
\end{equation}
Taking a double Laplacian, we then find the generalized Poisson equation:
\begin{equation}
\partial^4\phi(r) = -2n_{0,f}\beta\phi(r) + \partial^4\phi_{bare}(r)
\end{equation}
Taking the bare potential to be that of an isolated fracton, we obtain:
\begin{equation}
\partial^4\phi(r) = -2n_{0,f}\beta\phi(r) + Q\delta^{(3)}(r)
\end{equation}
Taking a Fourier transform, we have:
\begin{equation}
\begin{split}
k^4\phi(k) &= -2n_{0,f}\beta\phi(k) + Q \\
\phi(k) &= \frac{Q}{k^4 + 2n_{0,f}\beta}
\end{split}
\end{equation}
At large $k$ (short distances), the $k^{-4}$ behavior will lead to the same linear potential as the unscreened fracton, as expected.  However, at small $k$ we have that $\phi(k)$ approaches a constant.  This indicates that, in real space, $\phi(r)$ decays exponentially at long distances, with a characteristic screening length given by:
\begin{equation}
\lambda\propto (n_{0,f})^{-1/4} \propto e^{m_f/4T}
\end{equation} 
We see that, at the longest timescales, potentials and electric fields decay exponentially, leaving only short-range correlations in the system.  We refer to this as the strong-screening regime, in contrast with the power-law behavior of the weak-screening regime, brought about by dipole screening.

While all potentials and fields will eventually become exponentially screened, the timescale for this to happen can be quite large.  In order for fracton-on-fracton screening to become effective, we need fractons from the thermal bath to move distances of order the screening length $\lambda \propto (n_{0,f})^{-1/4}$.  As we identified in the previous section, the typical velocity of fractons will scale as $n_{0,d}$, arising from the absorption of dipoles.  The timescale for strong screening to come into effect then behaves as $\lambda/v \sim (n_{0,f})^{-1/4}(n_{0,d})^{-1}$.  (Note that the fracton is no longer undergoing a random walk, but rather being pulled in by the perturbing potential.)  As a function of temperature, the strong-screening timescale behaves as\cite{foot55}:
\begin{equation}
\tau_{ss}\propto (n_{0,f})^{-1/4}(n_{0,d})^{-1} \propto e^{(\frac{1}{4}m_f + m_d)/T}
\end{equation}
The important aspect of this formula is its exponential growth as $T$ is lowered.  At low temperatures, one would need to wait unreasonably long times for strong screening to come into effect, and the power-law behavior of weak screening will hold over experimentally relevant timescales.

This is an interesting intermediate regime which has no analogue in the conventional $U(1)$ spin liquid.  For the rank 1 case, as soon as a finite temperature is turned on, exponential screening quickly comes into effect and all correlations of observables decay exponentially.  For the fractonic $U(1)$ spin liquid described here, however, a charged disturbance will leave residual power-law correlations in the system for exponentially long times at low temperature.  These finite-temperature correlations will not be as long-ranged as the corresponding zero temperature ones.  Nevertheless, the existence of these long-range correlations, and their eventual disappearance at long times, may serve as an interesting way to diagnose the presence of a $U(1)$ fracton phase.

\subsection{Thermalization}

Our analysis has indicated that, at low temperatures, the motion of $U(1)$ fractons is still heavily suppressed, just as in the discrete fracton models\cite{abhinav}.  An interesting question which we can now ask is, starting from some fixed (non-thermal) configuration of fractons, how long does it take for these fractons to come to thermal equilibrium?  For example, starting from a stationary finite density $n$ of fractons, how long does it take for these fractons to relax to a thermal distribution?

Let us first ask this question in the presence of a thermal bath of dipoles, in which case each fracton will effectively perform a random walk, where each step takes a time proportional to $n_{0,d}^{-1}$.  We expect that, in order to thermalize, each fracton must travel at least a distance of order the interfracton spacing, $n^{-1/3}$, so that the fractons can mix together and smooth out the distribution.  By random walk physics, this requires about $n^{-2/3}$ steps.  We can therefore bound that the thermalization time of the system should be at least:
\begin{equation}
\tau_{therm} \sim n^{-2/3}n_{0,d}^{-1}\sim e^{m_d/T}
\end{equation}
which grows rapidly when the temperature becomes small.  Of course, this behavior will not persist to arbitrarily low temperatures.  When the temperature becomes low enough, dipoles will no longer be present in sufficient quantity to facilitate motion.  At the lowest temperatures, fractons will move primarily via motion mediated by other fractons.

As discussed in earlier work \cite{mach}, a fracton will acquire finite mobility in the presence of a finite-density distribution of other fractons.  In the presence of fracton density $n$, the typical velocity of a fracton will behave as $v\sim n$.  Once again, we expect that thermalization will only occur after fractons have moved a distance on the order of the interparticle spacing, $n^{-1/3}$.  In this case, we can bound that the thermalization time should be at least:
\begin{equation}
\tau_{therm}\sim \frac{n^{-1/3}}{n} \sim n^{-4/3}
\end{equation}
We therefore see that the thermalization time of an isolated fracton distribution ($i.e.$ without a thermal dipole bath) can also grow to be large when the fracton density is small, though the growth is now merely algebraic, instead of exponential.  But at low enough densities, such a state may still appear non-thermal on experimentally relevant timescales.  One is reminded of the physics of many-body localization.  Importantly, however, the resistance to thermalization of this system occurs even in the clean limit, without the need for any disorder.  This is an interesting feature of fracton models, which has been noted in the past \cite{chamon,abhinav}.  Before concluding this section, we note that there is one more process which might alter the above analysis: fracton decay.  We will discuss this issue in the Appendix, where we will find that fracton decay does not represent an important contribution to the thermalization of the system.

\section{Extensions to Other Higher Rank $U(1)$ Spin Liquids}

In the preceding sections, we have focused on a specific rank 2 $U(1)$ spin liquid phase with both fractons and nontrivial mobile excitations.  We found that this behavior leads to two distinct screening regimes.  At short times, only the mobile excitations can contribute to the screening, leaving weak power-law correlations in the system.  At much longer timescales, fractons will be able to fully screen each other, leaving only short-range correlations in the system.  For other higher rank $U(1)$ spin liquids with both fractons and mobile excitations, the story is likely much the same.  Mobile excitations will play a dominant role in screening on short timescales, but they will not be able to fully screen the long-range interactions.  On much longer timescales, however, we expect full screening to occur via the rearrangement of fractons.  Some of the scaling estimates determined in previous sections will likely be different, depending on the details of the phase, but the qualitative physics should be unchanged.

On the other hand, there are two other possible types of higher rank $U(1)$ phases, which we briefly mention in turn.

\subsection{``Type 2" $U(1)$ Fracton Models}

From earlier work on discrete fractons\cite{fracton2}, we know that there are certain models, including Haah's code \cite{haah}, in which all nontrivial excitations are fractons, whereas all mobile excitations are trivial ($i.e.$ can decay directly to the vacuum).  These models have been called ``type 2" fracton models.\cite{foot6}  To the best of the author's present knowledge, no one has yet explicitly written a $U(1)$ model which definitively has this property.  But little systematic exploration has been done on $U(1)$ models of rank higher than 2, and it seems likely that such ``type 2" models will be found somewhere in the higher spin hierarchy.

Assuming that such models exist, we expect them to have slightly different finite-temperature behavior.  Since there are no nontrivial mobile excitations, the system will have no screening ability at all on short timescales.  We therefore expect the full long-range correlations of the ground state to survive until some long timescale at which fracton-on-fracton screening sets in and makes everything short-ranged.  Such systems will therefore maintain a very strong ``memory" of their ground state behavior, even at finite temperature.

\subsection{Subdimensional Phases without Fractons}

There are also higher rank spin liquid phases which possess subdimensional particle excitations, but not fractons.  In a case like this, the fundamental charges have some degree of mobility, moving along a particular subspace, and can therefore rearrange their density to participate in screening.  Furthermore, the different subspaces are all coupled via the gapless gauge mode, ensuring that all subspaces come to the same thermal equilibrium.  We therefore expect phases like this will behave more like the conventional $U(1)$ spin liquid and will be fairly normal in their screening and thermalization properties.

As an example, let us consider a rank 2 $U(1)$ spin liquid with 1-dimensional excitations.  Our degrees of freedom will be the same as in the fracton model, a rank $2$ symmetric tensor gauge field $A_{ij}$.  The Gauss's law for this theory is:
\begin{equation}
\partial_i E^{ij} = \rho^j
\end{equation}
which defines a \emph{vector} charge density.  For the present purposes, the most important aspect of these charges is their conservation laws:
\begin{equation}
\int d^3r\,\rho^j = \textrm{constant}\,\,\,\,\,\,\,\,\,\,\,\,\,\,\,\,\,\,\,\int d^3r\,(\vec{r}\times\vec{\rho})^j = \textrm{constant}
\end{equation}
These equations represent the conservation of charge and the conservation of the angular moment of charge, respectively.  Unlike in the fracton model, the extra conservation law does not fully lock the charges in place, but rather constrains them to move only in the direction of their charge vector.  Thus, the charges in this theory are 1-dimensional particles, forced to live their lives along a line determined by their internal charge vector.

Let's work out precisely how screening works in this phase, again relying on some generalized electromagnetic results from Reference \onlinecite{genem}.  We have a set of vector charges, $\rho^j = \partial_i E^{ij}$, which have an electric field tensor $E_{ij}$ falling off as $1/r^2$, just as in conventional electromagnetism.  Unlike previous theories, we no longer have a scalar potential for describing electrostatic situations.  Instead, the potential formulation for this theory involves a vector-valued potential $\phi_i$, which gives the electric field via $E_{ij} = -\frac{1}{2}(\partial_i\phi_j + \partial_j\phi_i)$.  This vector potential essentially represents the potential energy per charge in a given direction.  The potential corresponding to a point charge $p^j$ takes the form:
\begin{equation}
\phi_p^j(r) = \frac{1}{8\pi}\bigg(\frac{(p\cdot r)r^j}{r^3}+3\frac{p^j}{r}\bigg)
\end{equation}
Let us now imagine a system at some finite temperature $T$.  We insert a test charge $P^j$ into our system, then see how the thermal distribution responds to the extra potential.  As in our earlier discussion of dipoles, the charge vectors of this theory are quantized in a lattice-dependent way.  (For example, on the cubic lattice, the charge vector is labeled by three integers, $\vec{p} = (n_x,n_y,n_z)$.)  As before, we only account for screening by the minimal charges, which is the dominant effect.  Let the thermally excited density of these minimal charges, in the absence of external perturbations, be denoted $n_0(T)\propto e^{-m/T}$, where $m$ is the mass of the minimal charges.  After we introduce a perturbing potential $\phi^j$, the density of species $p^j$ will become $n_0e^{-\beta p^j\phi_j}\approx n_0(1-\beta p^j\phi_j)$.  The total potential near the test charge will be:
\begin{equation}
\phi^j(r) = \phi^j_{bare}(r) - \sum_{p}\int dr'n_0\beta(p\cdot\phi(r'))\phi_p^j(r-r')
\end{equation}
where the sum is over the minimal charge vectors.  Using the fact that $-\frac{1}{2}(\partial^2\phi_p^j + \partial^j(\partial\cdot\phi_p)) = p^j\delta^{(3)}(r)$, we can then derive:
\begin{align}
\begin{split}
-\frac{1}{2}(\partial^2\phi^j + \partial^j(\partial\cdot\phi)) = -\frac{1}{2}&(\partial^2\phi_{bare}^j + \partial^j(\partial\cdot\phi_{bare})) \\ &- \sum_p n_0\beta(p\cdot\phi(r))p^j
\end{split}
\end{align}
As in our previous discussion of dipole screening, we can write $\sum_p p^ip^j = \alpha \delta^{ij}$ for some constant $\alpha$.  Also, assuming that the bare potential is that due to a point charge $P^j$, the equation becomes:
\begin{equation}
\partial^2\phi^j + \partial^j(\partial\cdot\phi) = -2P^j\delta^{(3)}(r) + 2 n_0\alpha\beta\phi^j(r)
\end{equation}
In momentum space, we then have:
\begin{equation}
-k^2\phi^j(k) -k^j(k\cdot\phi(k)) = -2P^j + 2n_0\alpha\beta\phi^j(k)
\end{equation}
We can rewrite this equation in tensor language as:
\begin{equation}
\bigg((k^2+2n_0\alpha\beta)\delta^{ij} + k^ik^j\bigg)\phi_i(k) = 2P^j
\end{equation}
Inverting for the potential, we obtain:
\begin{equation}
\phi^i(k) = \frac{2}{(k^2 + 2n_0\alpha\beta)}\bigg( \delta^{ij} - \frac{k^ik^j}{2(k^2 + n_0\alpha\beta)}\bigg)P^j
\end{equation}
The transverse component $\phi^j_\perp(k)$ (satisfying $k\cdot\phi_\perp = 0$) is given by:
\begin{equation}
\phi_\perp^j(k) = \frac{2P^j}{2n_0\alpha\beta + k^2}
\end{equation}
while the longitudinal component $\phi_\parallel$ (along the direction of $k$) is:
\begin{equation}
\phi_\parallel^j(k) = \frac{P^j}{n_0\alpha\beta + k^2}
\end{equation}
In both cases, $\phi(k)$ goes to a constant at $k=0$, and the potential will fall off exponentially in real space.  In either case, the screening length behaves as $\lambda\propto n_0^{-1}\propto e^{m/T}$.  Thus, at temperatures below $T\sim m/\log L$, spin liquid physics will hold throughout the system.

We see that the physics of screening in this phase closely mirrors that of the conventional $U(1)$ spin liquid, without anything particularly exotic.  Similar stories will hold for the other non-fractonic higher rank $U(1)$ spin liquids.  As long as the fundamental charges have some degree of mobility, they can quickly rearrange their density in such a way as to fully screen a test charge and kill all long-range interactions.  Since there is only a strong-screening regime, the physics of these spin liquids is not quite as rich as their fractonic counterparts.

\section{Conclusion}

In this paper, we have analyzed the finite-temperature screening behavior of higher rank $U(1)$ spin liquids, with particular emphasis on a phase with both fractons and nontrivial mobile excitations.  We have found that, as in their rank 1 counterparts, screening in these phases will allow for a smooth finite-temperature crossover to a trivial phase.  Nevertheless, the exponentially large screening length at low temperatures allows for a realistic range of temperatures at which spin liquid physics still governs the system.  Furthermore, for phases with fractons and mobile excitations, we have found that there are two distinct temporal regimes to the screening of charges.  On short timescales, a charged disturbance ($i.e.$ a fracton) is only weakly screened by the mobile excitations, leaving residual power-law correlations in the system.  At much longer times, fractons can screen each other, and all correlations become short-ranged.  The long timescales associated with screening and thermalization in these systems may be a useful tool for diagnosing the presence of such $U(1)$ fracton phases.

\section*{Acknowledgments}

I would like to acknowledge useful conversations with Senthil Todadri, Sagar Vijay, Rahul Nandkishore, Abhinav Prem, Claudio Chamon, Jonathan Ruhman, Liujun Zou, and Yahui Zhang.  This work was supported partially by NSF DMR-1305741 and partially by a Simons Investigator award to Senthil Todadri.

\section*{Appendix:  Fracton Decay}

In the main text, we estimated the time it takes for a system of fractons to move around and reach thermal equilibrium, both with and without a thermal bath of dipoles.  But how do we know that the fracton continues to exist at all over the relevant time scales?  When there are other fractons around, a fracton can recombine with an appropriate set of neutralizing charges to annihilate into the vacuum.  The energy gets carried away by the gapless gauge mode, which can freely propagate through the system, allowing for much faster thermalization.  Let us estimate the timescale for a fracton decay process.

First, we examine a state with a thermal distribution of dipoles.  Each fracton in the system is screened by a cloud of such dipoles.  As discussed in the main text, the composite object of a fracton plus its screening cloud is both charge-neutral and dipole-neutral, and can therefore decay directly into the vacuum.  However, at low temperatures, this object is large and extended, of characteristic size $\lambda$, the screening length.  From previous studies of the lattice models for these phases\cite{alex,sub}, we know that the matrix element for such a state to decay to the vacuum behaves as $e^{-(\lambda/a)^2}$, where $a$ is the lattice spacing.  The lifetime of the fractons then behaves as $e^{(\lambda/a)^2}\propto e^{(a^2n_{0,d}\beta)^{-1}}$, where the density of thermally excited dipoles is $n_{0,d}\propto e^{-m_d/T}$.  Therefore, fracton decay is enormously slow at low temperatures:
\begin{equation}
\tau_{decay} \sim e^{e^{m/T}}
\end{equation}
scaling as the exponential of an exponential.  Compared to the other thermalization processes discussed in the text, we find that fracton decay is too slow to play a significant role in the equilibration of the system.

Alternatively, let us start from the (non-thermal) state with fixed fracton density $n$, where the typical spacing between fractons is roughly $n^{-1/3}$.  If all the fractons have the same sign of charge, then there is no fracton decay mechanism available.  If there are a mix of positive and negative charges around, so that fracton decay can proceed, the lifetime of a fracton will be proportional to:
\begin{equation}
\tau_{decay}\sim e^{\frac{1}{(an^{1/3})^{2}}}
\end{equation}
At low densities, this timescale is much larger than the thermalization time due to fracton motion, which we found behaves as a power law in $n$.  Once again, fracton decay is not an important process contributing to thermalization.


\begin{thebibliography}{9}
\bibitem{foot1}
The positive and negative charges could in principle have different masses, but they will typically be of the same order.  We set these masses equal for simplicity.
\bibitem{foot2}
We will also in general allow the gauge field $A_{ij}$ to be compact, so that the charges are quantized.
\bibitem{foot3}
For the reader who is worried about a linearly growing potential, we refer to earlier work\cite{sub} on this ``electrostatic confinement" issue.  Even though the fractons in this model are energetically costly, they are still stable against conventional confinement.  Once the large energy cost is paid to create fractons, they are stable against collapsing back to the vacuum, due to their conservation laws.  Furthermore, not all $U(1)$ fracton models share this linear potential.  There exist more complicated fracton models which have normal decaying potentials, and to which the same screening principles apply.
\bibitem{foot4}
Note that in doing so we will pick up a boundary term, which does not in general vanish.  Nevertheless, after taking the Laplacian, this boundary term will not contribute to Equation \ref{dipolescreen}, so it is safe to ignore.
\bibitem{foot5}
Furthermore, the reduction to a $1/r^3$ electric field is already enough to eliminate the ``electrostatic confinement" issue mentioned earlier.
\bibitem{foot51}
The intuition for this exponential behavior is that motion of a composite object requires virtual dipoles to be exchanged across the object.  These dipoles are in general massive and have an exponentially decaying propagator, leading to exponentially decreasing hopping matrix elements as the composite object becomes larger.
\bibitem{foot55}
Note that this is a statistical average of the time necessary for a fracton to be screened, $i.e.$ a sort of ``mean-field" timescale.  Fractons in the tails of the statistical distributions could possibly take much longer to be screened.
\bibitem{foot6}
In Reference \onlinecite{fracton2}, fracton models could be characterized as ``type 1" or ``type 2" based on whether the fractons were created via membrane operators or via fractal operators.  In the $U(1)$ models, there are a richer variety of creation operators available, so this characterization is no longer strictly applicable.  Nevertheless, we can still distinguish type 1 versus type 2 models by whether or not they have nontrivial mobile excitations, which is a more robust characterization.
\end{thebibliography}
\end{document}